\documentclass[5p]{elsarticle}
\usepackage{hyperref}

\journal{Intermetallics}

\begin{document}

\begin{frontmatter}

\title{Anisotropic physical properties of PrRhAl$_4$Si$_2$ single crystal \\ a non-magnetic singlet ground state compound }


\author{Arvind Maurya}
\ead{arvind@tifr.res.in}
\author{R. Kulkarni}
\ead{ruta@tifr.res.in}
\author{A. Thamizhavel}
\ead{thamizh@tifr.res.in}
\author{S. K. Dhar}
\ead{sudesh@tifr.res.in}
\address{Department of Condensed Matter Physics and Materials Science,  Tata Institute of Fundamental Research \\
Dr. Homi Bhabha Road, Colaba, Mumbai 400 005, India}



\begin{abstract}
We have grown the single crystal of PrRhAl$_2$Si$_2$, which crystallizes in the tetragonal crystal structure.  From the low temperature physical property measurements like, magnetic susceptibility, magnetization, heat capacity and electrical resistivity, we found that this compound does not show any magnetic ordering down to 70~mK.  Our crystal field calculations on the magnetic susceptibility and specific heat measurements reveal that the 9-fold degenerate $(2J+1)$ levels of Pr atom in PrRhAl$_4$Si$_2$, splits into 7 levels, with  a singlet ground state and a well separated excited doublet state at 123~K, with a overall level splitting energy of 320~K.  
\end{abstract}


\end{frontmatter}


\section{Introduction}

The rare earth ion Pr based intermetallic compounds typically order magnetically, the exchange coupling between the localized Pr magnetic moments is provided by the indirect RKKY exchange interaction. Since Pr ($4f^2$) is a non-Kramer's ion, some compounds of Pr with a singlet crystal electric field (CEF) ground level show vanVleck paramagnetism and do not order magnetically. It is therefore of interest to explore the magnetic behaviour of new Pr compounds. The interest in Pr compounds is further heightened by the observation of heavy fermion superconductivity in PrOs$_4$Sb$_{12}$ in which the superconductivity has been attributed to a possible quadrupolar Kondo effect associated with a nonmagnetic $\Gamma_3$ doublet ground state~\cite{Bauer_PrOs4Sb12}. More recently, heavy fermion superconductivity in PrTi$_2$Al$_{20}$ under pressure in the vicinity of the quantum critical point of the quadrupolar order and heavy fermion superconductivity in the quadrupole ordered state of PrV$_2$Al$_{20}$ under ambient pressure has been observed~\cite{Sakai_PrTr2Al20}.  Recently, we have been investigating the anisotropic magnetic properties of RRhAl$_4$Si$_2$ (R =  Ce and Eu) and observed interesting magnetic behaviours in these compounds~\cite{Maurya_Structure_1142, Maurya_ETAS_magnetism, Maurya_CTAS_magnetism}.  In continuation to our studies on this type of compounds and to understand the crystal electric field level schemes in the non-Kramer's Pr-based intermetallic compound, in this communication we  report the magnetic properties of a new compound PrRhAl$_4$Si$_2$. The compound is isomorphic with the previously known RRhAl$_4$Si$_2$ (R~=~La, Ce and Eu) compounds~\cite{Maurya_Structure_1142}. Unlike the Ce and Eu compounds which order magnetically, due to the trivalent Ce-atom and the divalent Eu-atom, PrRhAl$_4$Si$_2$ is a vanVleck paramagnet; heat capacity measurements down to 70~mK do not reveal any evidence for a magnetic or quadrupolar transition. 

\section{Experimental}

Single crystals of PrRhAl$_4$Si$_2$ were grown by high temperature solution growth method using Al-Si eutectic as flux. The crystal was grown using the same protocol as already described for EuTAl$_4$Si$_2$ (T~=~Rh and Ir)~\cite{Maurya_Structure_1142}. In short, we have taken the stoichiometric ratio of high purity Pr, Rh, Al and Si elements in the ratio 1:1:4:2, together with the excess Al:Si eutectic (88:12) in a high quality recrystallized alumina crucible, which was subsequently sealed in a quartz ampoule under a vacuum of $10^{-6}$~Torr.   The ampoule was heated to 1050~$^\circ$C and held at this temperature for homogenization.  Then the furnace was slowly cooled at the rate of 2~$^\circ$C down  to 730~$^\circ$C  at which point the eutectic Al:Si flux was removed by centrifuging.  Several flat platelet like single crystals with very shiny metallic lustre, of dimensions of approximately $5~\times~5~\times~0.6~{\rm mm}^3$ were obtained.      Powder x-ray diffraction was used to check for phase purity, crystal symmetry and lattice parameters, while Laue diffraction was employed to establish the single crystal nature of the sample and to orient the crystal along the desired direction using a spark erosion cutting machine. Quantum design superconducting quantum interference device (QD SQUID) and Vibrating sample magnetometer (VSM) were used to measure the magnetization, while the data on heat capacity and electrical resistivity were taken on Quantum design physical properties measurement systems (QD PPMS).   The heat capacity of PrRhAl$_4$Si$_2$ was measured down to 70~mK using the dilution insert of the QD-PPMS.  

\section{Results and Discussion}

\subsection{Structure}
\begin{figure}[h]
\includegraphics[width=0.4\textwidth]{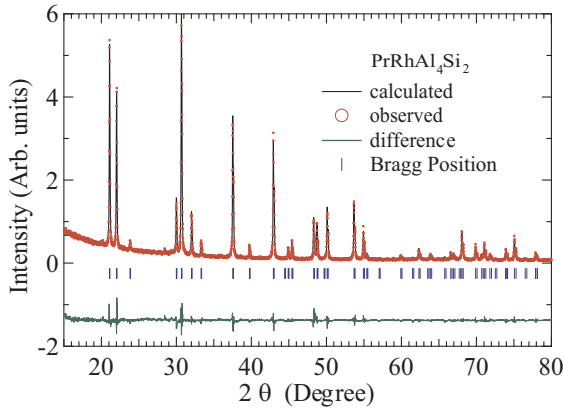}
\caption{\label{xrd}(Color online) Powder x-ray diffraction pattern of PrRhAl$_4$Si$_2$ and the Rietveld analysis. }
\end{figure}

Since the crystals of PrRhAl$_4$Si$_2$ were grown with off-stoichiometric ratio, the phase purity of the grown crystals was first confirmed by powder x-day diffraction.  The x-ray diffraction pattern of PrRhAl$_4$Si$_2$ along with the results of Rietveld refinement are shown in Fig.~\ref{xrd}.
The x-ray  diffraction pattern, is similar to that of RRhAl$_4$Si$_2$ (R~=~La, Ce and Eu) with a slight deviation in the peak position due to the difference in the lattice parameters.  From the Rietveld refinement it is confirmed that PrRhAl$_4$Si$_2$ crystallizes in the tetragonal crystal structure with the space group $P4/mmm$. The estimated lattice parameters are $a$~=~4.210(2)~\AA~and $c$~=~8.068(3)~\AA. The lattice parameters are slightly lower than the corresponding values in CeRhAl$_4$Si$_2$, which is in accordance with the lanthanide contraction.   Well defined Laue diffraction patterns, together with the four fold symmetry confirmed the tetragonal crystal structure.  The flat plane of the crystal has been identified as the (001) plane. 

\subsection{Magnetization}

\begin{figure}[h]
\includegraphics[width=0.4\textwidth]{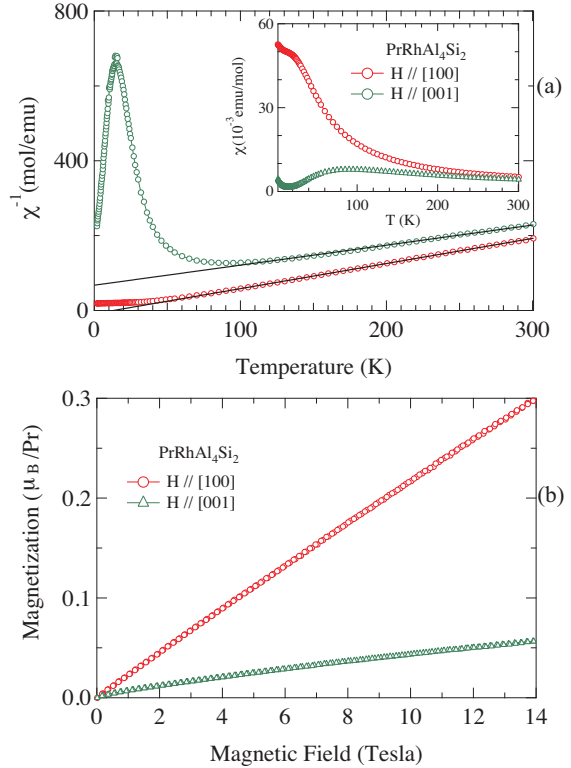}
\caption{\label{MT}(Color online)(a) Inverse magnetic susceptibility and susceptibility (in inset) up to 300~K of PrRhAl$_4$Si$_2$, (b) Isothermal magnetization measured at 2~K along the two principal crystallographic directions.}
\end{figure}

The inverse magnetic susceptibility, $\chi^{-1}$, measured in the temperature range from 1.8 to 300~K is shown in Fig.~\ref{MT}, while the susceptibility, $\chi$,  is plotted in the inset.  The susceptibility is anisotropic and a fit of Curie-Weiss expression to the inverse susceptibility from 150 to 300~K provides the following parameters:$\mu_{\rm eff}$~=~3.45 and 3.86~$\mu_{\rm B}$/Pr and $\theta_{\rm p}$~=~13 and -126.2 K for $H~\parallel$~[100] and [001], respectively.
The paramagnetic Weiss temperature $\theta_{\rm p}$  along [001] is almost an order of magnitude larger than its value along [100] and is of opposite sign, indicating  anisotropic magnetic behaviour of this compound in which the Pr ions form infinite sheets in square co-ordination in the $ab-$plane, stacked periodically along the c-axis by a repeat distance nearly twice as large as the separation in the $ab-$plane.  The  $\mu_{\rm eff}$ is comparable to the Hund's rule derived value for Pr$^{3+}$ magnetic moment. The prominent peak in $\chi^{-1}$ along [001] may be taken to be a signature of a magnetic transition but the plots of $\chi~ vs.~T$ do not suggest any magnetic ordering (heat capacity data described below do not provide any evidence for a long range magnetic transition). The $\chi$ along both the directions tends to attain a limiting value at low temperatures; the upturn below 10~K may partly be due to the presence of some paramagnetic impurities. Also, the inverse magnetic susceptibility deviates from the linearity below 100~K, revealing that the degenerate $J=4$ multiplet is split by the crystal electric field separation with a separation energy of about 100~K.  As mentioned in the beginning, a non-magnetic singlet ground state is possible in Pr compounds; a non-magnetic quadrupolar doublet can also give rise to a paramagnetic ground state.  We show latter from our crystal electric field (CEF) calculation that the anisotropy in the magnetic susceptibility is well explained and the ground state is a singlet  state.  

The magnetization at 2~K for $H~\parallel$~[100] is almost linear up to 14~T (Fig.~\ref{MT}(b)), reaching a value of only 0.3 $\mu_{\rm B}$/Pr, which is relatively low compared to the saturation moment, g$_J$J~=~3$\mu_{\rm B}$, of Pr. The data of Fig.~\ref{MT}(b) support a paramagnetic ground state of PrRhAl$_4$Si$_2$.
	 
\subsection{Heat capacity}
\begin{figure}[h]
\includegraphics[width=0.4\textwidth]{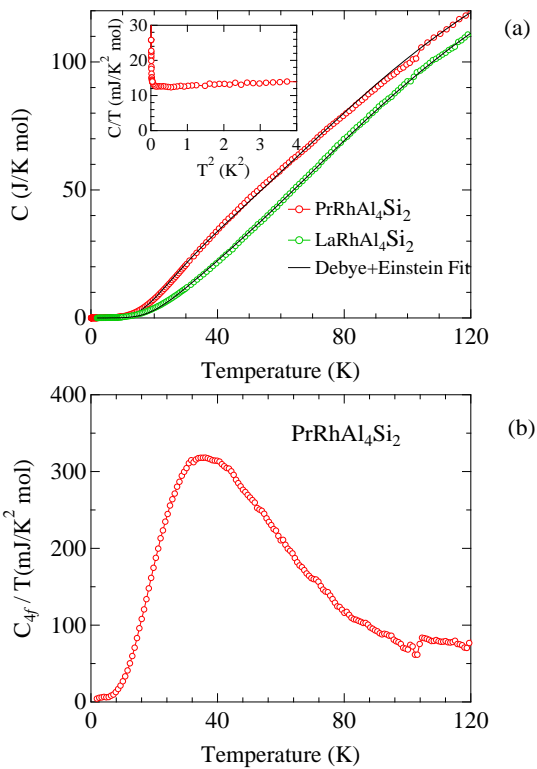}
\caption{\label{HC}(Color online) (a) Variation of heat capacity with temperature $C(T)$ of PrRhAl$_4$Si$_2$ and LaRhAl$_4$Si$_2$ down to 2~K. Inset shows the low temperature $C/T~vs~T^2$ data for sample PrRhAl$_4$Si$_2$ down to 70~mK.  The solid lines are the fits to the Debye plus Einstein model (refer text for details) (b) Magnetic part of the heat capacity obtained by subtracting the LaRhAl$_4$Si$_2$ from that PrRhAl$_4$Si$_2$.}
\end{figure}

The heat capacity of PrRhAl$_4$Si$_2$ between 1.8 and 150 K is shown in Fig.~\ref{HC}. The data for non-magnetic reference compound LaRhAl$_4$Si$_2$ are also plotted. The inset shows the heat capacity of the Pr compound at temperatures lower than 2 K as $C/T$ vs $T^2$ plot. The heat capacity does not show any evidence for a magnetic transition in PrRhAl$_4$Si$_2$ down to 70~mK. The upturn below 100~mK is due to the Schottky nuclear anomaly.  From the linear $C/T~vs~T^2$ plot we infer a value of $\sim$12~mJ/K$^2 \cdot {\rm mol}$ for the Sommerfeld coefficient.  The absence of any anomaly at low temperatures also rules out a transition associated with a possible non-magnetic quadrupolar doublet state, and indicates that the CEF ground level is a singlet. Such a conclusion is further supported by our CEF analysis described latter.

We have measured the heat capacity of LaRhAl$_4$Si$_2$ to deduce the magnetic part of the heat capacity of PrRhAl$_4$Si$_2$ and hence estimate the magnetic entropy.  The specific heat capacity did not approach to the Dulong and Petit's limit of $3nR$ (= 199.53~J/K mol) at 300~K.  This indicates that the Debye temperature for these compounds must be considerably higher than the room temperature.  An attempt to fit the specific heat data to Debye model, did not result in a good fit (not shown here for brevity).  Hence, we attempted to fit the heat capacity data to the following expression:

\begin{equation}
\label{Debye_Einstein}
C_{\rm p} =  [mC_{\rm Debye}(T) + (1-m) C_{\rm Einstein}(T)], 
\end{equation}

which includes the contribution from Debye and Einstein terms.  Where $C_{\rm Debye}$ and $C_{\rm Einstein}$ represents the usual expressions of the respective heat capacity which involves the Debye temperature ($\Theta_{\rm D}$) and the Einstein temperature ($\Theta_{\rm E}$).    Equation~\ref{Debye_Einstein} results in a best fit to the experimental data of PrRhAl$_4$Si$_2$  with 76\% of the weight to the Debye term with $\Theta_{\rm D} = 489~K$ and the remaining 24\% of weight to the Einstein term with $\Theta_{\rm E} = 106~K$, and for  LaRhAl$_4$Si$_2$  82\% of weight to $\Theta_{\rm D} = 499~K$ and the remaining 18\% to  $\Theta_{\rm E} = 131~K$.

The $4f$ contribution to the heat capacity, $C_{4f}$, derived by subtracting the heat capacity of LaRhAl$_4$Si$_2$ from that of PrRhAl$_4$Si$_2$, is plotted in Fig.~\ref{HC}(b). A prominent Schottky peak in $C_{4f}$ is centered around 50~K, and arises from the variation in the population of the CEF levels with temperature.  It may be noted that $C_{4f}$ shows a step near 100~K.  We believe it is due to the error in $C_{4f}$ which is derived by subtracting one large quantity from the other.  Besides, it may also be kept in mind that at high temperatures, the heat capacities of the two compounds may not be identical.  

An analysis of the experimental data of the magnetic susceptibility and the magnetic part of the heat capacity based on the CEF Hamiltonian is often helpful in providing an explanation of the observed behaviour in rare earth intermetallic compounds.  In order to proceed further with the CEF calculation, we need to know the rare-earth site symmetry in PrRhAl$_4$Si$_2$.  The Pr atom in PrRhAl$_4$Si$_2$ occupies the $1b$ Wyckoff's position of the $P4/mmm$ space group which possesses the tetragonal site symmetry $4/mmm$.  The CEF Hamiltonian for tetragonal symmetry is given as follows:

\begin{equation}
\label{CEF_eqn}
\mathcal{H}_{CEF} = B_2^0O_2^0 + B_4^0O_4^0 + B_4^4O_4^4 + B_6^0O_6^0+B_6^4O_6^6,
\end{equation}
  
where $B_l^m$ are the crystal field parameters and $O_l^m$ are the Steven's operators.  The Steven's operators are defined in Ref.~\cite{Hutchings, Stevens}.  The expression for the  magnetic susceptibility $\chi_{\rm i}$ calculated from the CEF Hamiltonian and including the molecular field contribution is given by
\begin{equation}
\label{eqn3}
\frac{1}{\chi_i} = \left(\frac{1}{\chi_{{\rm CEF}i}}\right)-\lambda_{i},  ~~~~(i = x, y, z)
\end{equation}

\begin{figure}[h]
\includegraphics[width=0.4\textwidth]{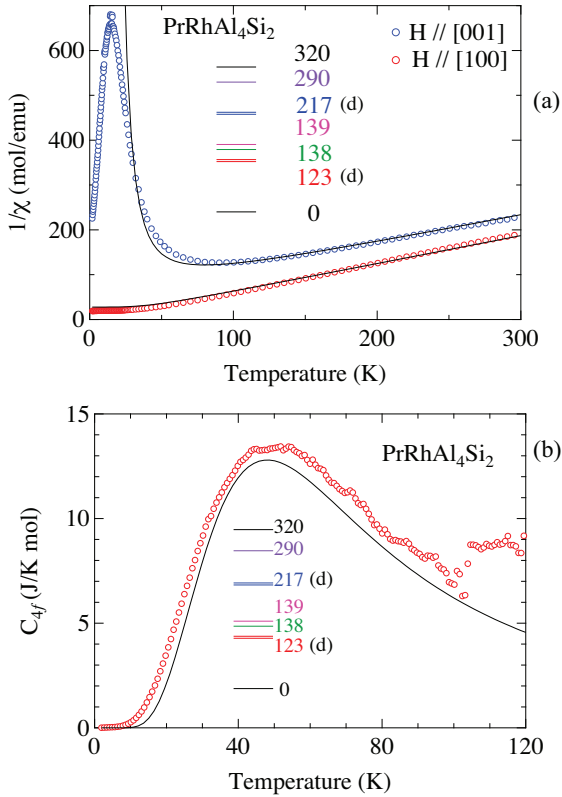}
\caption{\label{CEF}(Color online) (a) Inverse magnetic susceptibility plot together with the CEF calculated curve  along the two principal crystallographic directions, (b) magnetic part of the specific heat capacity and the calculated Schottky anomaly.  The obtained energy levels are also shown.}
\end{figure}

where $\lambda_{i}$ is the molecular exchange field constant and  $\chi_{{\rm CEF}i}$ is the combination of vanVleck term and the Curie term and is defined in Ref~\cite{Pranab_CeMg3}. We have obtained the crystal field parameters from the CEF calculations on the magnetic susceptibility data and the resultant $9~\times~9$ matrix is diagonalized to obtain the energy levels.  The total angular momentum for Pr is $J=4$ and hence there are  $2J+1$, 9-fold degenerate levels. The point charges surrounding this Pr atom lifts the 9-fold degeneracy.   Since the Pr atom, possesses the tetragonal site symmetry, the degenerate levels are split into 7 levels~\cite{Runciman}.  The calculated inverse magnetic susceptibility curve is shown in Fig.~\ref{CEF}(a).  The CEF calculations do not take into account possible exchange interaction between the ground and the excited CEF levels, whic his beyond the scope of our analysis. Although, a different set of crystal field parameters resulted in a comparable fit to the magnetic susceptibility, we have chosen only that set of parameters, which could explain the observed Schottky anomaly as shown in Fig.~\ref{CEF}(b).  The obtained crystal field parameters are $B_2^0 = 0.58$~K,  $B_4^0 = -0.08$~K,  $B_4^4 = -0.08$~K,  $B_6^0 =3 \times 10^{-3}$~K,  $B_6^4 = 3 \times 10^{-4}$~K and the molecular exchange field constants along [100] and [001] directions are $\lambda_{\rm [100]} = -2$~mol/emu and $\lambda_{\rm [001]} = -35$~mol/emu .  The energy levels are given in Fig.~\ref{CEF}. It is evident from the obtained energy levels that the ground state is well separated from the first excited doublet with an energy splitting of 123~K which is consistent with the nonmagnetic nature of PrRhAl$_4$Si$_2$.  Although our estimated crystal field energy levels explains the magnetic susceptibility semi-quantitatively and reproduces reasonably well the peak of the Schottky heat capacity $C_{4f}$,   a detailed inelastic neutron diffraction measurement is necessary to confirm these levels.

\subsection{Electrical Resistivity} 
\begin{figure}[h]
\includegraphics[width=0.4\textwidth]{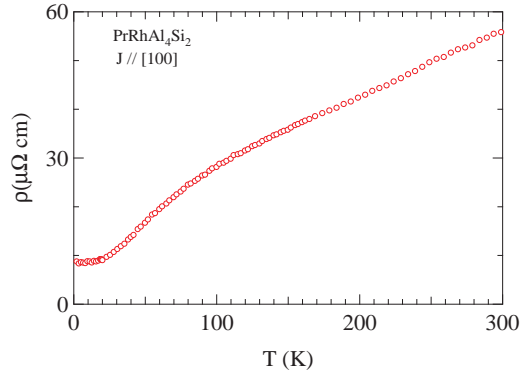}
\caption{\label{RT}(Color online) Electrical resistivity as a function of temperature when $J~\parallel$~[100] in PrRhAl$_4$Si$_2$ }
\end{figure}

The electrical resistivity of PrRhAl$_4$Si$_2$ as a function of temperature $\rho(T)$ for $J~\parallel~a$ is shown in Fig.~\ref{RT}. $\rho(T)$ decreases with temperature, saturating at $\sim$~8$\mu\Omega$~cm 
below  $\sim$~20~K. The low value of the residual resistivity, implies the good quality of the grown single crystal.  The resistivity is purely metallic in nature, however a broad hump centered around 100~K is observed.  This broad hump is mainly attributed the crystal field energy levels where the thermal population of the levels occur.  As observed in the magnetic susceptibility and the specific heat data, the resistivity also did not show any magnetic ordering down to 1.8~K confirming the nonmagnetic singlet ground state of PrRhAl$_4$Si$_2$.

\section{Conclusion}

We have successfully grown the single crystal of PrRhAl$_4$Si$_2$ by using the Al:Si eutectic composition as flux. Flat platelet like single crystal with the flat plane corresponding to (001) plane were obtained.  The grown crystals were oriented and cut along the principal crystallographic directions for the anisotorpic property measurements.  The sample is found to be highly anisotropic reflecting the tetragonal crystal structure of this compound.   All our magnetic measurements evinced the non magnetic nature of this compound.  The crystal field calculation performed on the magnetic susceptibility and the specific heat data clearly depict the singlet ground state of this compound.  

\section*{References}


\end{document}